\documentstyle[12pt,epsf]{article}
\newcommand{\be}{\begin{equation}}
\newcommand{\ee}{\end{equation}}
\newcommand{\al}{\mbox{$\alpha$}}

\newcommand{\bi}[1]{\bibitem{#1}}
\newcommand{\fr}[2]{\frac{#1}{#2}}

\newcommand{\gf}{\mbox{$\gamma_{5}$}}

\newcommand{\Ima}{\mbox{Im}}
\newcommand{\Rea}{\mbox{Re}}

\begin{document}
\pagestyle{empty}
\normalsize
\begin{flushright}{UQAM-PHE/96-09\\November 1996}
\end{flushright}
\vspace{0.5cm}
\begin{center}{\Large \bf $FCNC$ in left-right symmetric theories and constraints on the right-handed scale}\\

\vspace{1.5cm}

{\bf M.E.Pospelov }\footnote{E-mail:pospelov@mercure.phy.uqam.ca} \\

\vspace{0.5cm}

{\em 
D\'epartement de Physique, Universit\'e du Qu\'ebec \`a
Montr\'eal,\\ Case Postale 8888, Succ. Centre-Ville, Montr\'eal,
Qu\'ebec, Canada, H3C 3P8}\\
and\\
{\em Budker Institute of Nuclear Physics, 630090 Novosibirsk, Russia}

\end{center}
\vspace{1.5cm}
\begin{abstract}
We revise the limits on the $FCNC$ higgses in manifestly left-right symmetric theories. It is shown that the combination of the Kobayashi-Maskawa CP-violation with the tree level $\Delta S=2$ higgs exchange gives very large contribution to the CP-violating $\epsilon$ parameter. It leads to the new strong constraint on the $FCNC$ higgs mass, $M>50- 100$ TeV, enhanced almost by factor of the order $\sqrt{m_t/m_c}$. Being addressed to the supersymmetric left-right models, $FCNC$ problem requires both right-handed scale and supersymmetric mass parameters be heavier than $50$ TeV for $\tan\beta\sim 1$. The most relaxed case corresponds to $\tan\beta\sim 20- 30$ where right-handed scale can be of the order of few TeV. 
\end{abstract}
\newpage

\pagestyle{plain}
\pagenumbering{arabic}

\section{Introduction} 

Left-right symmetric extension of the Standard Model based on the group $SU(2)_L\times SU(2)_R \times U(1)_{B-L}$ have been the subject of theoretical studies for more than twenty years \cite{mohapatra}. The atractiveness of this model is in the possibility to have parity breaking as a result of the spontaneous breaking of the $SU(2)_R$ gauge group at some large mass scale $v_R$. The phenomenological consequencies of the additional right-handed gauge bosons in the kaon physics were studied in Refs. \cite{BBS,kaon,GE,Fr}. The LR box contribution to the $K^0-\bar{K^0}$ mixing is known to be enhanced in comparison with the SM case which provides the limit on the mass of the right-handed W-boson, $M_{W_{R}}>1.5$ TeV.  

However, the left-right symmetric model have one serious problem. Among the physical higgses there are two which violate flavour by two units and can bring unacceptably large contribution to the $K_L$-$K_S$ mass splitting \cite{kaon}. The corresponding lowest limit on the masses of these higgses in the left-right symmetric theories was found as $5- 10$ TeV. The common way to solve this problem is to relate these masses with the scale of the parity breaking $v_R$ \cite{GGMKO,DGKO}. 

In this article we study manifestly left-right symmetric theories in which the solution to the $FCNC$ problem has the simplest form. These theories have no additional sources of CP-violation like the complex phase in the left-right mixing. Therefore all CP-violating phenomena reside to the complex phase in the Kobayashi-Maskawa (KM) matrix. We show that the combination of the Kobayashi-Maskawa CP-violation with the exchange of the $FCNC$ higgs boson leads to very large contribution to the CP-violating $\epsilon$ parameter. This puts a new, very strong, limit on the masses of the $FCNC$ higgses and indirectly raises the lowest bound on the mass of the right-handed gauge boson. 

Then we consider the consequences of the $FCNC$ exchange in the left-right supersymmetric models. These models have been attracted a lot of attention recently. Besides some interesting phenomenological applications \cite{SUSYLR}, these models provide the natural suppression of the $CP$-violating $\theta$-parameter \cite{MR} and possess the explicit $R$-symmetry which could be broken only spontaneously \cite{MK,HM}. 

The simple solution for the $FCNC$ problem in the framework of the approach \cite{GGMKO,DGKO} does not work for supersymmetic models due to the more constrained form of the scalar potential in SUSY models. The terms responsible for the solution of $FCNC$ in usual left-right models simply do not exist here. We try to find the optimal choice of supersymmetric mass parameters and $v_R$ which would allow to have the most relaxed constraints coming from $FCNC$.

\section{The limit on the $FCNC$ higgs in LR model}

The minimal left-right symmetric model based on the $SU(2)_L\times SU(2)_R \times U(1)_{B-L}$ group has one scalar bidoublet $\phi$ with quantum numbers $(\fr{1}{2},\fr{1}{2}^*,0)$ and two triplets, $\Delta_L$ and $\Delta_R$ with quantum numbers (1,0,2) and (0,1,2) respectively:
\begin{eqnarray}
\Phi  = \left (\begin{array}{cc}
\phi^0_1&\phi^+_1\\\phi^-_2&\phi^0_2
\end{array}\right),
\Delta_{L}  = \left(\begin{array}{cc}
\Delta_L^+/\sqrt{2}&\Delta_L^{++}
\\\Delta_L^0&-\Delta_L^+/\sqrt{2}
\end{array}\right),
 \Delta_{R}  = \left(\begin{array}{cc}
\Delta_R^+/\sqrt{2}&\Delta_R^{++}
\\\Delta_R^0&-\Delta_R^+/\sqrt{2}
\end{array}\right).
\label{eq:fields}
\end{eqnarray}

The interaction of scalar fields with fermions has the following standard form:
\begin{eqnarray}
{\cal L}_Y=Y_u\bar{Q}_L \phi Q_R + Y_d \bar{Q}_L \tilde{\phi} Q_R\,
+Y_\nu\bar{E}_L \phi E_R + Y_e \bar{E}_L \tilde{\phi} E_R+\,H.c.;\\
{\cal L}_M=if(E_L^TC^{-1}\tau_2\Delta_LE_L+
E_R^TC^{-1}\tau_2\Delta_RE_R) + H.c.,\nonumber
\label{eq:yukawa}
\end{eqnarray}
where $\tilde{\phi}\equiv \tau_2\phi^*\tau_2$; $Q_{L(R)}$ and $E_{L(R)}$ are the quark and lepton doublets, $Y_i$ are corresponding yukawa couplings and $C$ is the charge conjugation matrix in Dirac space. Under the left right symmetry we have $\phi\leftrightarrow\phi^\dagger$, $\Delta_L\leftrightarrow\Delta_R$ which requires all $Y$-matrices to be hermitean in the generation space. $F$ matrix has to be symmetric.

Two stages of the $SU(2)_L\times SU(2)_R \times U(1)_{B-L}$ symmetry breaking are related with the vacuum expectation values (vevs) for neutral components in Eq. (\ref{eq:fields}):
\begin{eqnarray}
\langle \Phi \rangle = \left (\begin{array}{cc}
\fr{\kappa}{\sqrt{2}}&0\\0&\fr{\kappa'}{\sqrt{2}} e^{i\alpha}
\end{array}\right),
\langle \Delta_{L} \rangle = \left(\begin{array}{cc}
0&0\\v_{L}&0
\end{array}\right),
\langle \Delta_{R} \rangle = \left (\begin{array}{cc}
0&0\\v_{R}&0
\end{array}\right).
\nonumber
\end{eqnarray}

To satisfy the phenomenological requirements $v_{R}~\gg~max(\kappa,\kappa')~\gg~v_{L}$ at the very beginning we assume that $\Delta_L=0$. In this conditions the seesaw mechanism related with $ {\cal L}_M$-interaction always works and provides the suppression for the lightest neutrino mass eigenstate by the inverse power of $v_R$. From here and below we would concentrate our analysis on the quark sector.

The first line in Eq. (\ref{eq:yukawa}) gives masses to quarks but at the same time leads to $FCNC$ problem. Here we want to consider this problem in more details and, following the work \cite{GGMKO}, we assume for a moment that $\kappa'=0$. This case provides a natural way to to have one light higgs, curing the unitarity in the left-handed sector, and heavy $FCNC$ higgses associated with  the imaginary and the real part of the field $\phi^0_2\equiv(\phi^0_{2r}+i\phi^0_{2i})/\sqrt{2}$. The masses of gauge bosons and quarks in this case are written in the simplest form: 
\begin{eqnarray}
M_{W_R}^2=g^2v_R^2+\fr{g^2\kappa^2}{4};
M_{Z_2}^2\simeq 2(g^2+g_{B-L}^2)v_R^2;\nonumber\\
M_{W_L}^2=\fr{g^2\kappa^2}{4};
M_{Z_1}^2\simeq\fr{g^2\kappa^2}{4\cos^2\theta_W}\nonumber\\
M_u=Y_u\fr{\kappa}{\sqrt{2}};\; M_d=Y_d\fr{\kappa}{\sqrt{2}}.
\end{eqnarray}

The absence of CP-violation in the gauge sector leaves the  complex phase in the KM matrix $V$ to be the only source of CP-violation in the theory. (We do not consider here leptons where similar but independent phase may exist.) The hermiticity of the quark mass matrices ensures also the exact relations between mixing angles in the left- and right-handed sectors. 

The same yukawa term (\ref{eq:yukawa}) which gives masses to $U$-type of quarks, brings $FCNC$ interaction into the $D$-sector:
\be
{\cal L}_{FCNC}= 
\bar{D}_LV^\dagger Y^{diag}_uVD_R\phi_2^0+h.c.=
\bar{D}V^\dagger {\cal M}_uVD\fr{\phi_{2r}^0}{\kappa}+
i\bar{D}V^\dagger {\cal M}_uV\gf D\fr{\phi_{2i}^0}{\kappa}.
\label{eq:FCNC}
\ee
Here $Y^{diag}_u$ and ${\cal M}_u$ are the diagonal forms of the yukawa and mass matrices in the $U$-sector. 

The $d-s$ coupling in (\ref{eq:FCNC}) allows to have $\Delta S=2$ transition at the tree level. The corresponding $\Delta S=2$ Lagrangian reads as follows:
\be
{\cal L}_{\Delta S=2}=\fr{1}{M^2\kappa^2}\left(
\sum_{j=u,c,t}V^*_{jd}m_jV_{js}\right)^2
[(\bar{d}\gf s )^2 - (\bar{d} s)^2],
\ee
where we have assumed the common mass for the scalar and pseudoscalar higgs. 
After choosing KM matrix in the usual Wolfenstein parametrization 
\begin{eqnarray}
V=\left(\begin{array}{ccc}
1-\fr{\lambda^2}{2}&\lambda &A\lambda^3\sigma e^{-i\delta}\\
-\lambda &1-\fr{\lambda^2}{2}&A\lambda^2\\
A\lambda^3(1-\sigma e^{i\delta})&-A\lambda^2&1
\end{array}\right)
\end{eqnarray}
we are ready to calculate real and imaginary parts of the  $K^0-\bar{K^0}$ transition. To sufficient accuracy the results could be presented in the following simple form:
\begin{eqnarray}
\label{eq:Re}
\Rea\langle K^0|{\cal L}_{\Delta S=2}|\bar{K}^0\rangle=
\left(\fr{M_K}{m_s+m_d}\right)^2
\fr{f_K^2M_K^2}{\kappa^2M^2}
m_c^2\lambda^2\times\nonumber\\\left[ 
(1+\fr{m_t}{m_c}A^2\lambda^4(1-\sigma\cos\delta))^2
-\fr{m_t^2}{m_c^2}A^4\lambda^8\sigma^2\sin^2\delta\right]
\end{eqnarray}
\begin{eqnarray}
\Ima\langle K^0|{\cal L}_{\Delta S=2}|\bar{K}^0\rangle=
-\left(\fr{M_K}{m_s+m_d}\right)^2
\fr{f_K^2M_K^2}{\kappa^2M^2}
2m_cm_tA^2\sigma\lambda^6\sin\delta\times\nonumber\\
\left[ 
1+\fr{m_t}{m_c}A^2\lambda^4(1-\sigma\cos\delta)\right]
\label{eq:Im}
\end{eqnarray}
where we have used vacuum insertion method to calculate matrix elements. The quark masses in Eqs. (\ref{eq:Re})-(\ref{eq:Im}) are taken at the scale of $1$ GeV. (This is the only consequence of the one-loop QCD renormalization of the operator $\bar{q}_Lq_R\bar{q}_Rq_L$ \cite{GE}). 
The comparison of the Eq. (\ref{eq:Im}) with the standard model contribution (and with the experimental data) yields the following, very well known \cite{kaon}, constraint on the $FCNC$ higgs boson mass in LR theories:
\be
M>5- 10~\mbox{TeV},
\ee
where the main uncertainty comes from the undefinite relative sign between $m_t$ and $m_c$. It is easy to see also that the prediction for the imaginary part of the mixing is more certain than that for the real part. Moreover, the imaginary part is very large in this case and corresponding CP-violating parameter is of the order
\be
|\epsilon|\simeq
\fr{\Ima\langle K^0|{\cal L}_{\Delta S=2}|\bar{K}^0\rangle}
{2\sqrt{2}\Rea\langle K^0|
{\cal L}^{SM}_{\Delta S=2}|\bar{K}^0\rangle}
\simeq 2.0\cdot r\left(\fr{100\mbox{TeV}}{M}\right)^2
A^2\lambda^4\sigma\sin\delta\sim 1.5\cdot 10^{-3}r\left(\fr{100\mbox{TeV}}{M}\right)^2,
\ee
where parameter $r$ denotes the combination from the square brackets in Eq. (\ref{eq:Im}). To be in an agreement with the experimentally observed value of $\epsilon$, the mass of the $FCNC$ higgs has to be increased almost by the factor $\sqrt{m_t/m_c}$ so that the resulting bound is:
\be
M>50- 100~\mbox{TeV}
\label{eq:II}
\ee
The lower number, $50$TeV, corresponds to the case of the opposite signs for $m_t$ and $m_c$ and $r\sim 0.5$. If $m_c$ and $m_t$ are of the same sign the data on $CP$-violating $\epsilon$-parameter require $FCNC$ higgses be heavier than $100$ Tev. The bound (\ref{eq:II}) raises, though indirectly, the constraint on the right-handed gauge boson mass which cannot be lighter than $M/13$ \cite{OE} if we want the perturbative expansion to be preserved. 

The constraint (\ref{eq:II}) coming from the KM CP-violation mediated at the tree-level can be used to limit the ratio $\kappa'/\kappa$ which coincides with the admixture of the $FCNC$ scalar field into the higgs particle responsible for the unitarity in the left-handed sector (SM higgs):
\be
\fr{\kappa'}{\kappa}<(10^{-3}- 2\cdot 10^{-3}) \fr{m_{SM}}{100\mbox{GeV}}.
\label{eq:smh}
\ee

The constraints obtained above are held, in principle, for the case of nonvanishing CP-violating phase $\alpha$ in $W_L$-$W_R$ mixing if we do not assume accidental cancellations between contributions of different origin in $\epsilon$. The real alternative to the scenario outlined above is to have small $\sin\delta$ in the KM matrix  and lighter $FCNC$ higgs as a result. This represents, in some senses, the old-fashion superweak interaction scenario. To this end it is useful to rewrite the constraint (\ref{eq:II}) in the form of the relation:
\be
\fr{M}{\sqrt{|\sin\delta|}}\simeq 50- 100~\mbox{TeV}.
\label{eq:J}
\ee
The possibility of having small $\sin\delta$ was advocated in the recent work \cite{B}. In the model with small but nonvanishing ratio of $\kappa'/\kappa$ and symmetric real mass matrices the CP-violation originates from the relative phase $\alpha$ of two vevs. It turns out that there is a possibility to relate the CP-violating phase $\delta$ in 3 by 3 KM matrix with the parameter $\kappa'/\kappa\sin\alpha$ through the following relation \cite{B}:
\be
\sin\delta\simeq\fr{\kappa'}{\kappa}\sin\alpha
\fr{m_c}{m_s}\fr{s_2}{s_3}
\ee
Here $s_2$ and $s_3$ are the mixing angles between second and third and between first and third generations respectively. Combining it with our constraint (\ref{eq:J}) and taking the mass of the $FCNC$ higgs to be around the "old" limit of $10$ TeV, we obtain the tight bound on the spontaneous CP-violation in the model:
\be
\fr{\kappa'}{\kappa}\sin\alpha< 10^{-3}
\ee
In any case, the $FCNC$ higgs contribution to the kaon mixing is very important and cannot be neglected in comparison with the LL and LR box diagrams. 

To conclude this section we would like to remind the way how the $FCNC$ higgs can acquire the much larger mass than the left-handed electroweak scale \cite{GGMKO,DGKO}. The big degree of arbitrariness in the choice of the scalar potential allows, in principle, to relate the mass of the $FCNC$ higgs bosons with $v_R$. It can be achieved through the specific quartic couplings in the potential, 
${\cal L}_{scalar}=\alpha_1(\mbox{Tr}\Delta^\dagger_R\phi^\dagger\phi
\Delta_R+\mbox{Tr}\Delta^\dagger_L\phi^\dagger\phi
\Delta_L)+\alpha_2(\mbox{Tr}\Delta^\dagger_R\tilde{\phi}
^\dagger\tilde{\phi}
\Delta_R+\mbox{Tr}\Delta^\dagger_L\tilde{\phi}^\dagger\tilde{\phi}
\Delta_L)$, so that the $FCNC$ mass is proportional to the right-handed scale \cite{GGMKO}:
\be
M^2\simeq \fr{\alpha_1-\alpha_2}{2}v_R^2.
\label{eq:M-R}
\ee

\section{$FCNC$ problem and supersymmetric LR theories}

The stringent bound on the $FCNC$ higgs masses obtained in the previous section stimulates us to consider the same $FCNC$ problem in the framework of the supersymmetric left-right (SUSY LR) theories where the scale of the right-handed physics is usually assumed to be near $1$ TeV. 

The higgs content of the theory is two times larger than that of non-SUSY models. Now we have two bi-doublets, $\Phi_u$ and $\Phi_d$ with quantum numbers 
$(\frac{1}{2},\fr{1}{2}^*,0)$ and four triplets $\Delta_L(1,0,2)$, $\delta_L(1,0,-2)$, $\Delta_R(0,1,2)$ and $\delta_R(0,1,-2)$. The  yukawa interaction with quarks is the simple generalization of Eq. (\ref{eq:yukawa}):
\be
{\cal L}_Y=Y_u\bar{Q}_L \Phi_u Q_R + Y_d \bar{Q}_L \Phi_d Q_R\,+
\,H.c.
\ee
It leads to the same $FCNC$ interaction and same troubles in kaon sector. 

SUSY LR theories are known to suffer from the following serious problem. The scalar potential in SUSY LR models suggests that the vacuum configuration with $\langle \Delta_R \rangle = v_R \tau_1$ (and thus with the vev for doubly charged higgs) is lower than the charge conserving vacuum with $\langle \Delta_R \rangle = v_R \tau_-$ \cite{MK,HM}. The most natural way out of this problem is to assume that the other neutral scalar, sneutrino field $\tilde{\nu}$, develops vev which helps to avoid charge breaking minimum \cite{MK,HM}. The alternative way \cite{MR,MR1} is to hope that some remnants of the plank scale physics in the form of the effective operators dim=5 would shift the "wrong" vacuum up. Another very interesting feature of the SUSY LR models is the reality of vevs and the absence of spontaneous CP-violation \cite{MR,CP,MR1}. This implies that the manifest left-right symmetry is held, KM matrix is the only source of CP-violation and $\sin\delta\sim 1$. Therefore all constraints obtained in the previous section on the masses of $FCNC$ higgses can be generalized for the SUSY case.

We shall exploit here the approach to calculate the scalar mass spectrum developed in Ref. \cite{HM}. In this work the detailed analysis of the higgs mass spectrum was done and it was shown that it is possible to construct viable left-right supersymmetric theory using just one extra parameter $\langle \tilde{\nu}_R \rangle= \sigma_R$. Unfortunately, both mass spectra presented in \cite{HM}, with $\tan\beta=1.5 $ and $\tan\beta=50 $ give unacceptably large contribution to the mass  splitting and/or $\epsilon$-parameter in kaons.  In the first case ($\tan\beta=1.5 $) the contribution to the $\Delta m_{K_L-K_S}$ is about $1000$ times larger than the observed mass difference. The $\tan\beta=50 $ mass sample gives acceptable contribution to $\Delta m_{K_L-K_S}$ but violates $\epsilon$ constraint and also the $FCNC$ constraint coming from the $D^0-\bar{D}^0$ mixing. In what follows we try to find the optimal choice of the parameters which would allow SUSY LR to satisfy $FCNC$ constraints in the most relaxed way. 

The whole analysis of the scalar mass spectrum is very complicated because it requires the evaluation of the mass matrixes of the big dimension. Fortunately, the case $\kappa_u'=\kappa_d'=0$
\be
\langle\Phi_u\rangle = \left(\begin{array}{cc}\fr{\kappa_u}{\sqrt{2}}&0
\\0&0\end{array}\right);\;
\langle\Phi_d\rangle = \left(\begin{array}{cc}0&0
\\0&\fr{\kappa_d}{\sqrt{2}}\end{array}\right).
\ee
is simple in the sector of $FCNC$ higgses, which mass matrix  factorizes from the rest of scalars \cite{HM}. This choice is also strongly suggested by the $FCNC$ problem. Indeed, the supersymmetry requires at least one higgs particle to be light, of the order $M_Z$. The admixture of the flavour changing scalar fields in this particle is of the order $\kappa_u'/\kappa_u$. Then substituting $M_z$ instead of $m_{SM}$ in the constraint (\ref{eq:smh}) we find that this ratio has to be very small, $\kappa_u'/\kappa_u < 10^{-3}$, to avoid a huge contribution to $\epsilon$-parameter. In any case, we do not loose generality when setting $\kappa_1'$ and $\kappa_2'$ to zero because the spontaneous CP-violation is absent and $\sin\alpha=0$. 

The possibility to relate right-handed scale with the masses of $FCNC$ higgses through the relation (\ref{eq:M-R}) does not exist because the relevant couplings in SUSY LR are simply equal to zero, $\alpha_1=\alpha_2\equiv 0$. Therefore, the only way to create large masses for $FCNC$ higgses is to raise right-handed scale and all supersymmetric masses simultaneously. The mass matrix in the pseudoscalar $FCNC$ sector is the following:
\be
V=(\phi^{u\dagger}_{2i};
\phi^{d\dagger}_{1i})
\left(
\begin{array}{cc}
m^2\fr{\kappa_u}{\kappa_d}-D&m^2\\
m^2& m^2\fr{\kappa_d}{\kappa_u}+D,
\end{array}
\right)
\left(\begin{array}{c}
\phi^{u}_{2i}\\
\phi^{d}_{1i}
\end{array}\right),
\label{eq:mass}
\ee
where $D$ represent the contribution coming from the supersymmetric $D$-term:
\be
D=g_2^2\left(\fr{1}{2}\sigma_R^2+\langle\delta_R\rangle^2
-\langle\Delta_R\rangle^2+
\fr{\kappa^2_u}{2}-\fr{\kappa_d^2}{2}\right)
\ee
$m^2$ is the parameter from the soft breaking potential of the order of typical $m_{susy}$:
\be
V=...+m^2\mbox{Tr}[\tau_2\Phi_d^T\tau_2\Phi_u].
\ee
The contribution of the $F$-term is neglected in (\ref{eq:mass}) because it is proportional to the square of yukawa couplings which we regard to be small. Two mass eigenstates of the mass matrix (\ref{eq:mass}) are 
\be
\fr{m_{1,2}^2}{m^2}\simeq \fr{\tan\beta+\cot\beta}{2}
\pm\sqrt{\left(\fr{\tan\beta+\cot\beta}{2}\right)^2+
\left(\fr{D}{m^2}\right)^2-\fr{D}{m^2}(\tan\beta-\cot\beta)}
\ee
where $\tan\beta=\kappa_u/\kappa_d$. Varying three parameters, $m$, $D$ and $\tan\beta$, we want $m_1$ and $m_2$ to satisfy all $FCNC$ constraints.

\begin{flushleft}{\em Case 1.} $\tan\beta\sim {\cal O}(1)$\end{flushleft}

If $\tan\beta\sim {\cal O}(1)$ the mixing angle between  $\phi^{u}_{2i}$ and $\phi^{d}_{1i}$ is large and both pseudoscalars have to obey the limit (\ref{eq:II}). It is possible only if $m^2\sim D^2 \sim (50- 100\mbox{TeV})^2$. The mass of $W_R$ cannot be smaller than $\sqrt{D}$ and therefore $M_{W_R}$ is pushed to a very high scale:
\be
M_{W_R}^2= g_2^2\left(\fr{1}{2}\sigma_R^2+\langle\delta_R\rangle^2
+\langle\Delta_R\rangle^2+
\fr{1}{4}(\kappa^2_u+\kappa_d^2)\right)>D>(50~\mbox{TeV})^2
\ee
It should be noted also that in the approach of Refs. \cite{MR,MR1} $M_{W_R}^2\gg D$.

\begin{flushleft}{\em Case 2.} $\tan\beta\gg 1$\end{flushleft}

When $\tan\beta$ is large the mixing angle between $\phi^{u}_{2i}$ and $\phi^{d}_{1i}$ is small and proportional to $\cot\beta$. In this case one pseudoscalar almost coincides with $\phi^{u}_{2i}$ and is much heavier than the other one. 
\begin{eqnarray}
m_1^2\simeq m^2\tan\beta\\\nonumber
m_2^2\simeq D
\label{eq:c}
\end{eqnarray}
The first mass is constrained again by (\ref{eq:II}) so that the limit on the supersymmetric mass is:
\be
m_{susy}\sim m > \fr{50- 100 \mbox{TeV}}{\sqrt{\tan\beta}}
\label{eq:c1}
\ee
 The second particle causes problems in the kaon sector only through the small admixture of $\phi^{u}_{2i}$-field. The resulting constraint on $D$ is 
\be
\sqrt{D}=m_2>\fr{50- 100 \mbox{TeV}}{\tan\beta}
\label{eq:c2}
\ee
From the other hand, large $\tan\beta$ leads to the large yukawa couplings in the $D$-type of quarks and therefore $\phi^{d}_{1i}$ is capable to produce significant $D^0-\bar{D}^0$ mixing through the diagram of the Fig. 2. The corresponding constraint reads as follows \cite{LNS}:
\be
m_2>\tan\beta\cdot 100 \mbox{GeV}.
\label{eq:c3}
\ee
Combining together all three constraints (\ref{eq:c1})-(\ref{eq:c3}) we obtain the most relaxed choice of parameters in SUSY LR with respect to $FCNC$ problem:
\begin{eqnarray}
\tan\beta\sim 20- 30\nonumber\\
M_{W_R}>\sqrt{D}>2.5- 5~\mbox{TeV}\\
m_{susy}> 10- 20~\mbox{TeV}\nonumber
\end{eqnarray}
Our analysis shows that the soft breaking masses still have to be very large but at the same time $M_{W_R}$ can be of the order of few TeV. 

\section{Strong CP-problem and $FCNC$}

The tight relation between these two problems was noticed in \cite{MR1}. Here we would like to present very important radiatively induced contribution to $\bar{\theta}$, closely related with $FCNC$, and find correct combination of the yukawa couplings and mixing angles which gives the biggest contribution to $\theta$. 

In our recent work \cite{P} we have shown that the large $\tan\beta$ leads to the large radiative corrections to $\bar{\theta}$-parameter due to the squark-gluino loop correction to the quark mass:
\be
\bar{\theta}\sim \Ima(V^*_{td}V_{tb}V^*_{cb}V_{cd})y_c^2y_t^4\fr{m_b}{m_d}
\fr{\al_s}{4\pi}
\fr{3\al_w}{2\pi}\mbox{ln}\fr{M^2_{W_R}}{M^2_{W_L}}
\fr{m_{\tilde{G}}(A-\mu\tan\beta)}{m_{squark}^2}
\sim 10^{-9}\tan\beta
\label{eq:t1}
\ee
CP-violation here emerges through the multiple yukawa matrix insertions into the squark line. These insertions is the result of one-loop renormalization from $GUT$ scale down to $M_{W_L}$. The logarithmic factor $\mbox{ln}(M^2_{W_R}/M^2_{W_L})$ arises through the difference $m_R^2-m_L^2$ between masses of right- and left-handed squarks. 

If $\tan\beta\sim {\cal O}(1)$ the theory does not need axion-related mechanism for the relaxation of $\theta$. The result (\ref{eq:t1}) was obtained under the assumption that the low-energy theory has MSSM structure which implies also the following relations between different scales in the problem: $M_{W_L}\sim m_{susy}\ll M_{W_R},~M_{FCNC}$. This condition, however, is very artificial and cannot be achieved automatically from the superpotential. To this end it is reasonable to investigate the  different case of $M_{W_L}\ll m_{susy}\sim M_{W_R}\sim M_{FCNC}$.

We have to consider another group of diagrams, Fig. 3, which can give potentially large contribution to $\bar{\theta}$ proportional just to the first power of the charm quark yukawa coupling. The relevant part of the quark-squark-higgsino interaction, the complete analogue of the interaction (\ref{eq:FCNC}) involving superpartners, looks as follows:
\be
{\cal L}_{FCNC}= 
\bar{D}_R\tilde{\phi}_2^uV^\dagger Y^{diag}_uV\tilde{D}_L+
\tilde{D}^*_RV^\dagger Y^{diag}_uV\bar{\tilde{\phi}_2^u}D_L~+h.c.
\ee
$CP$-violation in this diagram arises due to one power of $y_t^2$-insertion into the $D$-squark line where $y_t$ is the top quark yukawa coupling. It does not bring any additional suppression because $y_t\sim 1$. The condition $m_{susy}\sim M_{W_R}$ implies also that $m_R^2-m_L^2$ cannot be obtained as a result of the logarithmic evolution and simple one-loop insertion into the squark line. Instead, we have to consider whole set of two-loop corrections and dress the diagram of the Fig. 3 in all possible ways by one loop involving $W_L$ or $W_R$. To get a crude estimate we use the result (\ref{eq:t1}) where we change couplings and put logarithm to one:
\be
\bar{\theta}\simeq\fr{\Ima~m_d}{\Rea~m_d}\sim \Ima(V^*_{td}V_{tb}V^*_{cb}V_{cd})y_cy_t^3\fr{m_b}{m_d}
\fr{1}{16\pi^2}\fr{3\al_w}{2\pi}
\fr{m_{\tilde{\phi}}(A-\mu\tan\beta)}{m_{squark}^2}.
\label{eq:t2}
\ee
The scalar mass $m_{squark}$ is presumably of the same order as $m_{\tilde{\phi}}$. Numerically this expression constitutes:
\be
\bar{\theta}\sim 10^{-8}\tan\beta~~\mbox{for}~\tan\beta>1.
\label{eq:t3}
\ee

The main numerical difference of two answers, (\ref{eq:t1}) and (\ref{eq:t2}), is in the smaller degree of suppression by the charm quark yukawa coupling and in the absence of the large logarithmic factor in (\ref{eq:t2}). Taken literally, this value of $\bar{\theta}$ (\ref{eq:t2}) leads to unacceptably large electric dipole moment of the neutron. We have to admit, however, that the estimate (\ref{eq:t3}) is not very far from experimental limit. It means that the more accurate calculation of $\bar{\theta}$ is required.

\section{Conclusions}

We have presented the arguments showing the importance of the $FCNC$ problem for the left-right symmetric models. The constraints coming from the CP-violating $\epsilon$-parameter are very strong, pushing the masses of $FCNC$ higgses close to $100$ TeV range. Being indirectly related with $M_{W_R}$ mass, this suggests that the scale of the right-handed gauge bosons is of the order $10$ TeV if we want to avoid strongly interacting right-handed physics. 

It is instructive to compare $FCNC$ problem in multi-higgs doublet models with that of left-right models. We have demonstrated that this problem is more sound in the LR models because $FCNC$ arises here in the down quark sector with the yukawa couplings coming from the $U$-type of quarks. The relations $y_c\gg y_s$, $y_t\gg y_b$ explain also why the constraint (\ref{eq:II}) derived here is much stronger than those usually obtained in multi-higgs models. At the same time we found much less freedom to escape this constraint in the manner acceptable for multi-higgs models \cite{WHWW} because the number of yukawa couplings in LR is twice less than in multi-higgs models if the number of the higgs doublets is the same. 

The same problem persists in the SUSY LR models. Here we do not have such a freedom in the scalar potential as it was in the usual LR. The only way to satisfy $FCNC$ bounds is to raise the supersymmetric mass parameter and at the same time $M_{W_R}$ scale. For $\tan\beta\sim 1$ the supersymmetric mass parameter and $M_{W_R}$, both have to be heavier than $50$ TeV. In this situation the left-handed scale itself is the result of the 1\%-accuracy fine tuning between large mass parameters from the soft breaking potential: $m_{\phi_u}^2|\Phi_u|^2 + m_{\phi_d}^2|\Phi_d|^2 + m^2\mbox{Tr}[\tau_2\Phi_d^T\tau_2\Phi_u]$.  We found, however, the possibility of relatively low $M_{W_R}$, of the order of few TeV, for the case of $\tan\beta\simeq 20- 30$.

We consider also $FCNC$-related contributions to $\bar{\theta}$-parameter relevant for the solution of strong CP-problem in left-right supersymmetric models. It is shown that the radiative correction to $\bar{\theta}$ is large and its correct dependence of the quark masses and mixing angle is $\Ima(V^*_{td}V_{tb}V^*_{cb}V_{cd})
y_cy_t^3\fr{m_b}{m_d} \sim 10^{-4}$ and even two-loop effects can introduce large $\bar{\theta}$.

Very high scale of the right-handed gauge bosons diminishes someone's hope to discover them at the next generation of colliders. The positive signal of the left-right symmetry would be the discovery of the doubly charged particles which are uneffected by $FCNC$ problem and still could be relatively light. 
 
\begin{flushleft}{\large {\bf Acknowledgements}}\end{flushleft}
I would like to thank G. Couture, M. Frank, C. Hamzaoui and H. K\"onig for many helpful discussions. This work is supported by NATO Science Fellowship, N.S.E.R.C., grant \#  189 630 and Russian Foundation for Basic Research, grant \# 95-02-04436-a.

\newpage
{\bf Figure captions}

Fig. 1 Tree-level $\Delta S=2 $ higgs exchange 

Fig. 2 Tree-level $\Delta C=2 $ higgs exchange

Fig. 3 $FCNC$-related diagram leading to the $\theta$-term in SUSY LR. 

\vspace{1cm}

\begin{figure}[hbtp]
\begin{center}
\mbox{\epsfxsize=144mm\epsffile{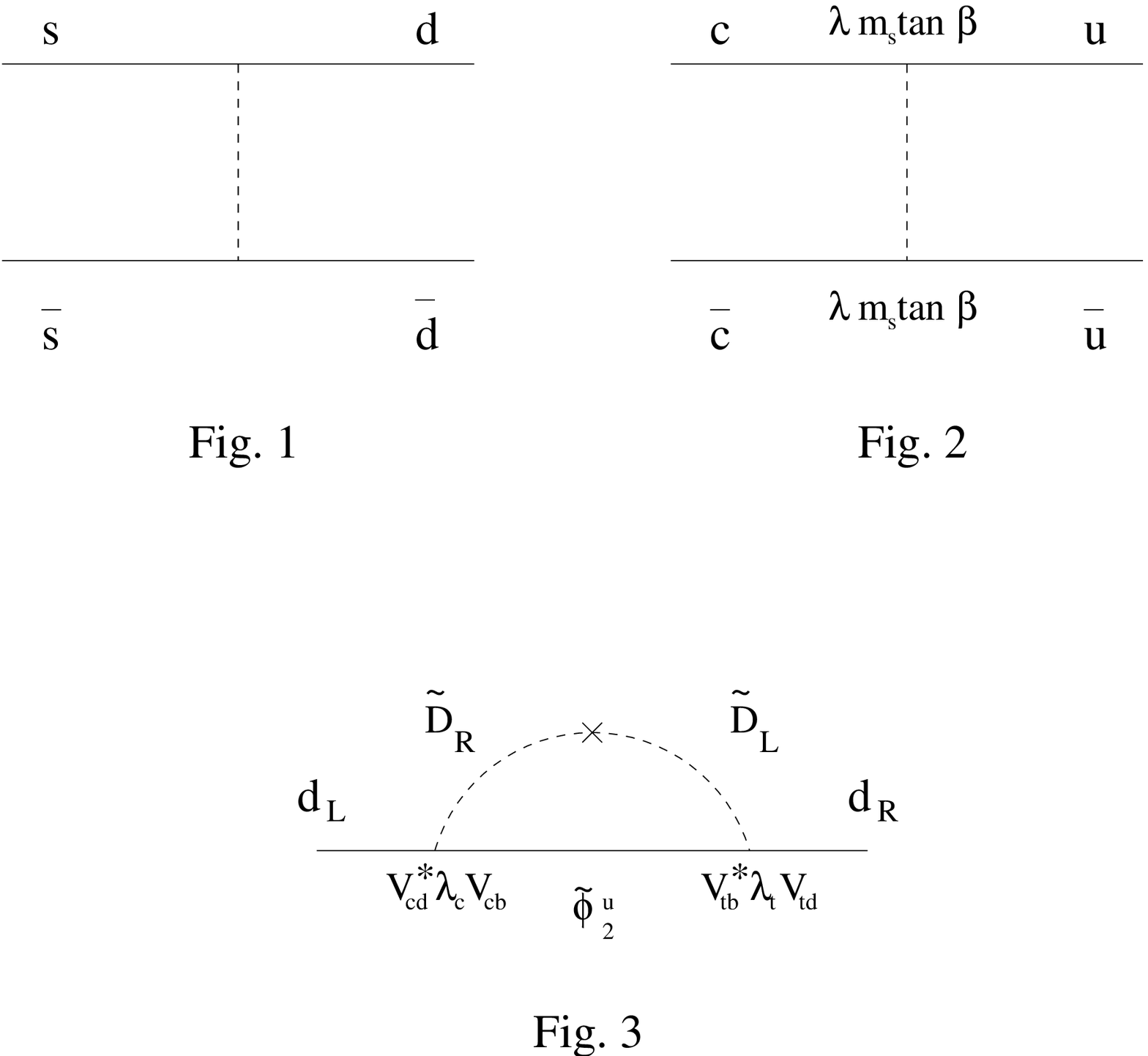}}
\end{center}
\label{muegf}
\end{figure}

\end{document}